\documentclass[laps,pra,twocolumn,floatfix,showpacs]{revtex4}

\usepackage[T1]{fontenc}
\usepackage{mathptmx}
\usepackage[mathscr]{eucal}
\usepackage{amsfonts}
\usepackage[colorlinks=true,citecolor=blue,linkcolor=blue,urlcolor=blue]{hyperref}
\usepackage{amsmath,amssymb,amsthm}
\usepackage{color}
\usepackage{graphics}
\usepackage[dvips]{graphicx}
\usepackage{psfrag}
\include{Qcircuit}

\begin{document}

\title{Unifying parameter estimation and the Deutsch-Jozsa algorithm for continuous variables}
\author{Marcin Zwierz}\email{php07mz@sheffield.ac.uk}
\author{Carlos A. P\'{e}rez-Delgado}\email{c.perez@sheffield.ac.uk}
\author{Pieter Kok}\email{p.kok@sheffield.ac.uk}
\affiliation{Department of Physics and Astronomy, The University of
Sheffield, Hounsfield Road, Sheffield, S3 7RH, United Kingdom}

\date{\today}

\begin{abstract}\noindent
We reveal a close relationship between quantum metrology and the Deutsch-Jozsa algorithm on continuous variable quantum systems. We develop a general procedure, characterized by two parameters, that unifies parameter estimation and the Deutsch-Jozsa algorithm. Depending on which parameter we keep constant, the procedure implements either the parameter estimation protocol or the Deutsch-Jozsa algorithm. The parameter estimation part of the procedure attains the Heisenberg limit and is therefore optimal. Due to the use of approximate normalizable  continuous variable eigenstates the Deutsch-Jozsa algorithm is probabilistic. The procedure estimates a value of an unknown parameter and solves the Deutsch-Jozsa problem without the use of any entanglement.
\end{abstract}

\maketitle

\section{Introduction}
\noindent
Quantum metrology promises many advances in science and technology. Continuous variables (CV), i.e. eigenstates of an operator $\hat{x}$ with continuous spectrum, are natural candidates for optical implementations of quantum metrology protocols \cite{caves,braunstein98,kok_book}. The importance of continuous variables for quantum metrology stems from the unconditional and efficient character of CV preparation, manipulation and detection techniques \cite{braunstein05,lloyd}. In this paper, we devise an optimal parameter estimation procedure for continuous variables. Our procedure employs a single continuous variable and estimates a value of an unknown parameter with Heisenberg-limited precision. Furthermore, for a particular, fixed value of the parameter in question the procedure behaves as the Deutsch-Jozsa algorithm for CVs. In fact our protocol extends the Deutsch-Jozsa algorithm over continuous variables presented by Pati and Braunstein \cite{pati}. Instead of idealized, nonnormalizable and unphysical states, we employ Gaussian states to represent continuous variables. Moreover, we define Gaussian states on a finite domain, thus removing an unphysical, infinite speed-up over any classical procedure offered by the idealized states. An extensive analysis of the Deutsch-Jozsa algorithm over continuous variables was given by Adcock, H{\o}yer, and Sanders \cite{sanders09}.

The Deutsch-Jozsa algorithm is one of the first quantum algorithms, preceded only by the original Deutsch algorithm \cite{deutsch}. Even though the Deutsch-Jozsa problem is rather artificial, the algorithm drew enormous attention due to the computational speed-up over any classical procedure. The structure of the algorithm is simple enough to determine the source of this speed-up. The quantum superposition principle and consequent quantum parallelism that lie at the heart of quantum mechanics allows for the interference of many distinct computational paths, and allows the correct answer to the problem emerge in a single query. In other words, the Deutsch-Jozsa algorithm probes a global property of an unknown function $f(x)$ and returns the result in a single run.

The paper is organized as follows. In Sec.~\ref{sec:dj}, we review the Deutsch-Jozsa algorithm over continuous variables and present its simplified version. In Sec.~\ref{sec:qm}, we review the basic concepts in quantum metrology. In Sec.~\ref{sec:unify}, we introduce a general procedure that unifies parameter estimation with the Deutsch-Jozsa algorithm, and analyze it in detail. Finally, we make some concluding remarks in Sec.~\ref{sec:conc}.

\section{Deutsch-Jozsa algorithm over continuous variables}\label{sec:dj}

\noindent
The generalization of the Deutsch-Jozsa algorithm to continuous variables was devised by Pati and Braunstein \cite{pati}. This generalization was implemented with idealized continuous variables defined on an infinite domain. However, we need to stress that any practical continuous-variable implementation of the Deutsch-Jozsa problem can be realized only on a finite domain. Nevertheless, for simplicity, we first recall the Deutsch-Jozsa algorithm over continuous variables as originally stated in Ref.~\cite{pati}.

The object of the Deutsch-Jozsa problem is to determine whether some function $f(x)$ is constant or balanced. This is achieved by Alice and Bob playing the following game. Alice submits a value of $x$ from $-\infty$ to $+\infty$ to Bob. Then Bob evaluates $f(x)$, which can take only two values: 0 or 1. Bob also promises Alice to use either balanced or constant functions. A constant function is either 0 or 1 for all values of $x \in (-\infty,+\infty)$. A balanced function is 0 for half of the values of $x$, and 1 for the remaining values of $x$. This is defined in terms of the Lebesgue measure $\mu$ on $\mathbb{R}$: $\mu ({x \in \mathbb{R}|f(x) = 0})=\mu ({x \in \mathbb{R}|f(x) = 1})$ \cite{pati}. The goal of this game is the same as the objective of the Deutsch-Jozsa problem, i.e. to establish if the function used by Bob was constant or balanced. Classically, Alice would have to submit infinitely many values of $x$ to learn the global property of $f(x)$ with certainty. However, if Bob can use a unitary black-box operation to calculate function $f(x)$, then only a single function evaluation reveals the global property of $f(x)$. In the setting of idealized continuous variables this would imply an infinite speed-up over any classical procedure.

\begin{figure}[t]
\includegraphics[width=6cm]{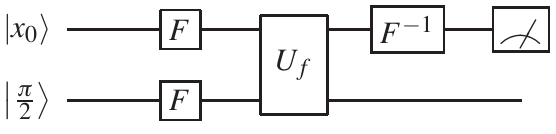}
\caption{A quantum circuit representing the Deutsch-Jozsa algorithm over continuous variables. The quantum network ${N_{DJ}}$ consists of the Fourier transforms $F$ and controlled black-box gate $U_{f}$ applied to the register and target CVs prepared in the idealized position eigenstates $|x_{0}\rangle$ and $|\pi/2\rangle$, respectively. The last operation is an inverse Fourier transform that enables the interference of different computational paths. \label{DJ_2CV}}
\end{figure}

Let us now introduce in some detail the ideal Deutsch-Jozsa algorithm over continuous variables shown in Fig.~\ref{DJ_2CV}. This implementation of the Deutsch-Jozsa algorithm employs two continuous variables, the so-called register and target CVs. Alice stores her query in the register CV and the target CV is used by Bob during function evaluation. The register CV is prepared in the position eigenstate $|x_{0}\rangle$ and the target in the eigenstate $|\pi/2\rangle$. The quantum network ${N_{DJ}}$ implementing the Deutsch-Jozsa algorithm is given by the following unitary transformation
\begin{equation}
N_{DJ} = F^{-1}_{r} U_{f} F_{r} F_{t}\, ,
\end{equation}
where $F$ is the Fourier transform and $r$, $t$ indicates the register and target CV, respectively. The Fourier transform applied to a CV in some position eigenstate $|x\rangle$ creates a superposition of all position eigenstates according to
\begin{equation}
F|x\rangle = \frac{1}{\sqrt{\pi}} \int^{\infty}_{-\infty} dy \, e^{2 i x y} |y\rangle,
\end{equation}
where we used photon number units in which $\hbar=\frac12$. The unitary black-box operator $U_{f}$ evaluates a value of function $f(x)$ and stores it in the state of the target CV: $|x \rangle |y \rangle \xrightarrow{} |x \rangle |y + f(x) \rangle$. Let us analyze the Deutsch-Jozsa algorithm step by step: (i) prepare the register and target CVs in an ideal position eigenstate $|x_{0}\rangle$ and $|\pi/2\rangle$, respectively; (ii) apply the Fourier transform $F$ to the register and target CVs
\begin{equation}
|s\rangle = F_{r} F_{t} |x_{0} \rangle |\pi/2 \rangle = \frac{1}{\pi} \int^{\infty}_{-\infty} dx dy \, e^{2i x x_{0}+ i \pi y} |x\rangle |y \rangle; \nonumber
\end{equation}
(iii) following the action of a unitary black-box operator $U_{f}$ the state of the CVs is given by
\begin{equation}
U_{f}|s\rangle = \frac{1}{\sqrt{\pi}} \int^{\infty}_{-\infty} dx \, e^{2i x x_{0}} e^{-i \pi f(x)}  |x\rangle F_{t}|\pi/2\rangle; \nonumber
\end{equation}
(iv) the quantum network ${N_{DJ}}$ is finalized with an inverse Fourier transform $F^{-1}$ applied to the register CV. Therefore, the state of the CVs can be written as
\begin{eqnarray}
F^{-1}_{r} U_{f} |s\rangle = \frac{1}{\pi}\int^{\infty}_{-\infty} dx dx' \, e^{2i x (x_{0}-x')} e^{-i \pi f(x)}  |x' \rangle F_{t}|\pi/2 \rangle; \nonumber
\end{eqnarray}
(v) following the quantum network ${N_{DJ}}$, the property of function $f(x)$ is determined by projecting the state of the register CV onto the original position eigenstate $|x_{0}\rangle$. The continuous-variable projection operator for idealized states can be written as
\begin{equation}
P_{x_{0}} = \int^{x_{0}+\varepsilon}_{x_{0}-\varepsilon} dy \, |y\rangle \langle y|,
\end{equation}
where $\varepsilon$ is the spread around $x_{0}$ value, i.e. the continuous-variable measurement cannot be performed with infinite precision. The orthogonal complement of $P_{x_{0}}$ is given by
\begin{equation}
P_{\bar{x}_{0}} = I - P_{x_{0}} = I - \int^{x_{0}+\varepsilon}_{x_{0}-\varepsilon} dy |y\rangle \langle y|.
\end{equation}
By construction, a complete set of orthogonal projectors $P_m$ satisfy the completeness relation $\sum_{m} P_{m} = I$ and $P_{m}P_{m'} = \delta_{m m'}P_{m}$. If $f(x)$  is constant then the measurement statistics based on the above set of orthogonal projection operators and assuming $\varepsilon \rightarrow 0$ is given by
\begin{eqnarray}
p(x_{0}) &=& \mbox{Tr}[\hat{P}_{x_{0}} \rho_{DJ}] = 1, \\
p(\bar{x}_{0}) &=& \mbox{Tr}[\hat{P}_{\bar{x}_{0}} \rho_{DJ}] = 0,
\end{eqnarray}
where $p(x_{0})$ is the probability of measurement outcome to be $x_{0}$, $p(\bar{x}_{0})$ is the probability of a measurement outcome different than $x_{0}$ and $\rho_{DJ} = N_{DJ} |r\rangle |t\rangle \langle t| \langle r| N^{-1}_{DJ}$. Conversely, if $f(x)$ is balanced then the measurement statistics assuming $\varepsilon \rightarrow 0$ is given by
\begin{eqnarray}
p(x_{0}) &=& \mbox{Tr}[\hat{P}_{x_{0}} \rho_{DJ}] = 0, \\
p(\bar{x}_{0}) &=& \mbox{Tr}[\hat{P}_{\bar{x}_{0}} \rho_{DJ}] = 1.
\end{eqnarray}
Therefore, if the state of the register CV remains unchanged then the function $f(x)$ is definitely constant, and if the state of the register CV is not $|x_{0}\rangle$ then the function $f(x)$ is balanced. A single function evaluation solves the Deutsch-Jozsa problem.

The core of the above implementation of the Deutsch-Jozsa algorithm is represented by a unitary, controlled black-box operator $U_{f}$ applied between the Fourier transformed register and target CVs. Here, the Fourier transformed target CV together with a black-box operator induces a phase shift, which depends on the global property of the function $f(x)$: $U_{f}(|x\rangle F_{t} |\pi/2\rangle) = e^{-2i f(\hat{x}) \hat{p}_{t}} |x\rangle F_{t}|\pi/2\rangle = e^{-i \pi f(x)} |x\rangle F_{t} |\pi/2\rangle$. Notice that the state of the target CV is not changed following the action of $U_{f}$. In fact $F_{t} |\pi/2\rangle$ is an eigenstate of $U_{f}$ with an eigenvalue $e^{-i \pi f(x)}$ ``kicked back'' in front of the register CV \cite{cleve}. Conventionally, the Deutsch-Jozsa algorithm employs multiple quantum systems, however, as the above simple analysis of the action of $U_{f}$ indicates the target CV can be omitted. It is easy to show that a single register CV together with a redefined black-box operator $U_{f} \equiv e^{-2 i \, \pi/2 \, f(\hat{x})}$ is enough to implement the Deutsch-Jozsa algorithm over continuous variables. In Ref.~\cite{sanders09}, the authors  arrived at the same conclusion, however, they used a slightly different approach. We emphasize that a direct consequence of employing a single system is that this protocol does not use any entanglement to determine the global property of the function in a single run. Moreover, the above implementation of the Deutsch-Jozsa algorithm is expressed in terms of the idealized position eigenstates. However, a more realistic and physically meaningful representation of a continuous variable is given by, for example, Gaussian states.

Similar to the setting of discrete quantum systems (e.g. qubits), some features of the Deutsch-Jozsa algorithm can serve as a starting point for developing other quantum algorithms. A slightly modified black-box operator $U_{f} \equiv e^{-2 i \, \pi/2 \, f(\hat{x})}$ for a simplified Deutsch-Jozsa algorithm can be used as the core of a protocol capable of estimating an unknown parameter that under appropriate conditions still retains the capabilities of the Deutsch-Jozsa algorithm. Before introducing this protocol, let us recall some basic concepts in quantum parameter estimation theory.

\section{Parameter estimation}\label{sec:qm}

\begin{figure}[!t]
\centering
\includegraphics[width=6cm]{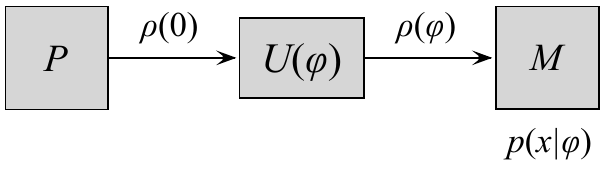}
\caption{The general parameter estimation procedure involving state preparation $P$, evolution $U(\varphi)$ and generalized measurement $M$ with outcomes $x$, which produces a probability distribution $p(x|\varphi)$. \label{PE}}
\end{figure}

\noindent
The most general parameter estimation procedure is shown in Fig.~\ref{PE}, and consists of three elementary steps: (i) prepare a probe system in an initial quantum state $\rho(0)$, (ii) evolve it to a state $\rho(\varphi)$ by a unitary evolution $U(\varphi)=\exp(-i\varphi\mathcal{H})$, (iii) subject the probe system to a generalized measurement $M$, described by a Positive Operator Valued Measure ({\sc povm}) that consists of elements $\hat{E}_x$, where $x$ denotes the measurement outcome. Here, the Hermitian operator $\mathcal{H}$ is the generator of translations in $\varphi$, the parameter we wish to estimate. The amount of information about $\varphi$ that can be extracted by a measurement of the probe system is given by the Fisher information
\begin{equation}
F(\varphi) = \sum_{x} \, \frac{1}{p(x|\varphi)} \left( \frac{\partial p(x|\varphi)}{\partial \varphi} \right)^2\, ,
\end{equation}
where $p(x|\varphi) = \mbox{Tr}[\hat{E}_{x} \rho(\varphi)]$ is the probability distribution given by the Born rule that describes the measurement data, and $x$ is a discrete measurement outcome.
Based on the Fisher information one can bound a minimal value of the uncertainty in $\varphi$ with the quantum Cram\'{e}r-Rao bound \cite{helstrom67,helstrom69,braunstein94}
\begin{equation}
(\delta\varphi)^2 \geq \frac{1}{T F(\varphi)}\, ,
\end{equation}
where $(\delta\varphi)^2$ is the mean squared error in the parameter $\varphi$, and $T$ is the number of times the procedure is repeated. The ultimate limit of the quantum Cram\'{e}r-Rao bound depends on how the Fisher information is bounded from above. The Fisher information can be bounded in two ways: by the variance of $\mathcal{H}$ \cite{braunstein96} or by the expectation value of $\mathcal{H}$ \cite{zwierz10,jones10}
\begin{equation}
F(\varphi) \leq 16(\Delta \mathcal{H})^2  \quad\mbox{and}\quad  F(\varphi) \leq 4 \langle\mathcal{H}\rangle^2, \label{bounds}
\end{equation}
where we again used $\hbar = \frac12$. Since both bounds are completely general and complement each other, any parameter estimation procedure must respect them. Typically, the Fisher information is related to an appropriate resource count such as the average photon number, the average energy of the probe system or the number of unitary evolution gates that are used in the estimation procedure. The expectation value of $\mathcal{H}$ plays the role of the resource count \cite{zwierz10}. We usually consider two scaling regimes of the quantum Cram\'{e}r-Rao bound. The first regime: the so-called {\em standard quantum limit} ({\sc sql}) \cite{gardiner04} or {\em shot-noise limit} is obtained when the Fisher information is a constant with respect to $T$ and the resource count. The {\sc sql} is typically given by
\begin{equation}
\delta\varphi \gtrsim \frac{1}{\sqrt{T}}\, .
\end{equation}
The second regime: the so-called {\em Heisenberg limit} \cite{holland93} is obtained in a single-shot experiment ($T=1$) when the Fisher information scales quadratically with the resource count. The Heisenberg limit is then given by
\begin{equation}
\delta\varphi \geq \frac{1}{\sqrt{F(\varphi)}}\, .\label{HL}
\end{equation}
Therefore, the uncertainty in the parameter $\varphi$ scales linearly inversely with the resource count. Both scaling regimes of the quantum Cram\'{e}r-Rao bound can be compared directly in terms of an appropriate resource count \cite{zwierz10}.

\section{General procedure with Gaussian states} \label{sec:unify}
\begin{figure}[!t]
\includegraphics[width=7.5cm]{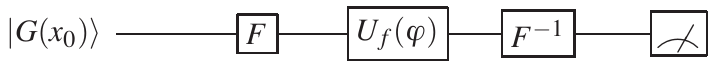}
\caption{A quantum circuit representing the general protocol over continuous variables. The quantum network consists of the Fourier transform $F$ and black-box gate $U_{f}(\varphi)$ applied to a single register CV prepared in the Gaussian state $|G(x_{0})\rangle$. The last operation is an inverse Fourier transformation that enables the interference of different computational paths. \label{estimation}}
\end{figure}

\noindent
In this section, we present a general procedure capable of determining the value of a single parameter $\varphi \in [0,2\pi)$ or implementing the Deutsch-Jozsa algorithm (see Fig.~\ref{estimation}).
Here, the black-box operator is defined in the following way
\begin{equation}
U_{f}(\varphi) \equiv \exp(-2i \varphi f(\hat{x}))\, ,
\end{equation}
where $f(\hat{x}) |x\rangle = f(x) |x\rangle$. The function $f(x)$ again takes only two values 0 and 1. Without loss of generality, ideal, nonnormalizable continuous variable states are regularized to Gaussian input states. Similar to the case of the Deutsch-Jozsa algorithm, any physical continuous-variables parameter estimation protocol can be implemented only on a finite domain. Therefore, we introduce the semi-Gaussian input state defined on a finite domain given by
\begin{equation}
|G(x_{0})\rangle = \int^{T}_{-T} \frac{dx}{N_{x}} \, \exp\left[-\frac{(x - x_{0})^2}{2 \Delta^2}\right] |x\rangle,
\end{equation}
where $\Delta$ is the variance of the state and $N_{x}$ is the normalization constant given by $N_{x}^{2} = \sqrt{\pi \Delta^2}/2 \, \left[\mbox{erf}(\frac{T + x_{0}}{\Delta}) + \mbox{erf}(\frac{T - x_{0}}{\Delta})\right]$. We note that for $\Delta \ll T$ we recover the normalization constant in the form of $N_{x}^2 = \sqrt{\pi \Delta^2}$ which is characteristic for a Gaussian state defined on an infinite domain, i.e. from $-\infty$ to $+\infty$. The Fourier transformed semi-Gaussian state defined on a finite domain can be written as
\begin{equation}
|G(p_{0})\rangle = \int^{P}_{-P} \frac{dp}{N_{p}} \, \exp\left[- 2 \Delta^2 (p - p_{0})^2\right] |p\rangle,
\end{equation}
where $1/(2 \Delta)$ is the variance of the Fourier transformed semi-Gaussian state and $N_{p}$ is given by
\begin{equation}
N_{p}^{2} = \frac{\sqrt{\pi/ 4 \Delta^2}}{2} \, \left[\mbox{erf}(2 (P + p_{0}) \Delta) + \mbox{erf}(2 (P - p_{0}) \Delta)\right]\, .\nonumber
\end{equation}
For $P \gg 1/(2 \Delta)$ the normalization constant takes the form of $N_{p}^2 = \sqrt{\pi/ 4 \Delta^2}$, characteristic for a Fourier transformed Gaussian state define on an infinite domain. The relationship between domains of the semi-Gaussian input state and its Fourier transformed counterpart is given by $P = 1/(2T)$.

The general procedure consists of the following instructions: (i) prepare the register CV in the normalized semi-Gaussian state $|r\rangle = |G(x_{0})\rangle$, and apply the Fourier transform $F$ defined by
\begin{equation}
F|x\rangle = |x\rangle_{p} =  \frac{1}{\sqrt{2T}} \int^{T}_{-T} dy \, e^{2 i x y} |y\rangle \, ,
\end{equation}
where $|x\rangle_{p}$ is the Fourier transformed position eigenstate, i.e. the momentum eigenstate; (ii) subsequently, a black-box operator $U_{f}(\varphi)$ is applied. Then the state of the system is
\begin{eqnarray}
U_{f}(\varphi) F|r\rangle &=& \int^{T}_{-T} \frac{dx}{N_{x}} \, \exp\left[{-\frac{\left(x - x_{0}\right)^2}{2 \Delta^2}}\right] e^{-2i \varphi f(\hat{x})} |x\rangle_{p} \nonumber \\
&=& \frac{1}{\sqrt{2 T}} \int^{T}_{-T} \frac{dx dy}{N_{x}} \, \exp\left[{-\frac{\left(x - x_{0}\right)^2}{2 \Delta^2}}\right] \nonumber \\ && \times e^{2i y x} e^{-2i \varphi f(y)} |y\rangle \nonumber \, ;
\end{eqnarray}
(iii) finally, an inverse Fourier transform $F^{-1}$ is applied followed by a measurement. The state of the register CV is measured by projecting onto the original semi-Gaussian state centered around $x_{0}$. Measurement is described by a POVM set $\{P_{x_{0}}, P_{\bar{x}_{0}}\}$, where
\begin{equation}
P_{x_{0}} = \int^{T}_{-T} dx dy \, g_{xy} |x\rangle \langle y|, \; \mbox{and} \; P_{\bar{x}_{0}} = \mathbb{I} - P_{x_{0}} \label{POVM}
\end{equation}
with
\begin{equation}
g_{xy} = \frac{1}{N_{\varepsilon}^{2}} \, \exp\left[-\frac{\left(x - x_{0}\right)^2}{2 \varepsilon^2}\right] \exp\left[-\frac{\left(y - x_{0}\right)^2}{2 \varepsilon^2}\right],
\end{equation}
and $\varepsilon$ is the intrinsic precision of the measurement apparatus, i.e. any continuous-variable measurement must have finite precision if it is to be physical, and $N_{\varepsilon}$ is the normalization constant given by $N_{\varepsilon}^{2} = \sqrt{\pi \varepsilon^2}/2 \, \left[\mbox{erf}(\frac{T + x_{0}}{\varepsilon}) + \mbox{erf}(\frac{T - x_{0}}{\varepsilon})\right]$. The optimal measurement which corresponds to the initial semi-Gaussian register state has $\varepsilon = \Delta$, thus $N_{\varepsilon} = N_{x}$.

Now let us calculate the measurement statistics. Analytical expressions for the measurement statistics are hard to find due to the presence of error functions $\mbox{erf}(x)$. However, for the semi-Gaussian states with $\Delta \ll T$ the calculations simplify considerably. Under this regime, the limits of integration for the integrals containing terms that depend on $\Delta$ range from $-\infty$ to $+\infty$. Necessarily, the normalization constants have to be changed and are expressed as $\sqrt{2T} N_{x} = \sqrt{\pi} \sqrt[4]{\pi \Delta^2}$. In other words, a semi-Gaussian input state defined on a finite domain is approximated with a Gaussian state defined on an infinite domain. Therefore, the measurement statistics based on the above POVM is given by the following expression
\begin{eqnarray}
p(x_{0}|\varphi) &=& \frac{4 \Delta^2}{\pi} \int^{P}_{-P} dz dy \, e^{-4 \Delta^2 (z^2 + y^2)} e^{2i \varphi (f(z)-f(y))}, \nonumber \\
p(\bar{x}_{0}|\varphi) &=& 1 - p(x_{0}|\varphi). \label{pro}
\end{eqnarray}
Here, the interval $(-P, P)$ is a finite domain of the Fourier transformed semi-Gaussian state $|G(x_{0})\rangle$, and denotes the interval, where for this particular procedure function $f(x)$ is defined.

At this point, we have to give an explicit definition of the function. Functions $f(x)$ defined on a finite domain returning only two values $\left(\{0,1\}\right)$ fall into three distinct categories: constant, balanced and neither constant nor balanced. We recall that the objective of the Deutsch-Jozsa algorithm is to probe whether an unknown function $f(x)$ is constant or balanced. We parameterize the three possibilities for defining $f(x)$ by introducing a parameter $r$. The above integrals can then be evaluated for any function $f(x)$ behaving as a step function, with the parameter $r$ marking the point where $f(x)$ changes its value. Hence, for $r = 0$ and $r = \pm P$ the function $f(x)$ is balanced and constant, respectively. For $0 < r < P$ (or $-P < r < 0$), the function $f(x)$ is neither constant nor balanced. We consider only positive values of $r$ due to the symmetry of the setup. This leads to
\begin{eqnarray}
p(x_{0}|\varphi) &=& \frac{1}{2} \left[\mbox{erf}^{2}(2 P \Delta) + \mbox{erf}^{2}(2 r \Delta)\right] + \nonumber \\
&& \frac{1}{2} \left[\mbox{erf}^{2}(2 P \Delta) - \mbox{erf}^{2}(2 r \Delta)\right]\cos(2 \varphi), \nonumber \\
p(\bar{x}_{0}|\varphi) &=& 1 - p(x_{0}|\varphi), \nonumber
\end{eqnarray}
where $p(x_{0}|\varphi)$ is the probability of measurement outcome to be in the interval $x_{0} \pm \varepsilon$ and $p(\bar{x}_{0}|\varphi)$ is the probability of measurement outcome not to be in the interval $x_{0} \pm \varepsilon$.

\subsection{Representations of $f(x)$ }\noindent
Our choice to represent $f(x)$ as a step function simplified our calculations. However, we can imagine more elaborate behavior patterns for $f(x)$. In principle, since in the case of the Fourier transformed idealized continuous variables all terms have amplitudes of equal magnitude, all finite sub-intervals, where the function takes value 0 can be added up to a single interval. The same applies to all sub-intervals, where function takes value 1. Therefore, one ends up with two intervals and a relationship between them given by the parameter $r$. However, in the setting of semi-Gaussian states defined on a finite domain, the above reasoning is not quite as straightforward. The amplitudes of the Fourier transformed Gaussian states have a slightly different magnitude. One may notice this feature by inspecting Eq.~(\ref{pro}). Since in our calculations we favor a step-function representation over any other, let us estimate the maximum error we make with this assumption. Due to a trivial nature of a constant function, in the following analysis we consider a balanced function. We consider the step-function representation of a balanced function with $r = 0$. The biggest deviation from this representation is offered by a balanced function that changes its value twice at points $r_{1} = -P/2$ and $r_{2} = P/2$.  Both representations produce two distinct probability distributions $p_{step}(x_{0}|\varphi)$ and $p_{hat}(x_{0}|\varphi)$, respectively, that differ by the error $\epsilon_{P\Delta}$ given by
\begin{equation}
\epsilon_{P\Delta} = \left|1 - \cos(2 \varphi)\right| \times \left|- \frac{8}{\pi} (P \Delta)^6 + \frac{24}{\pi} (P \Delta)^8 + O\left((P \Delta)^{10}\right)\right| \, . \nonumber
\end{equation}
The error tends to zero with $P \Delta \rightarrow 0$. This is natural since when $\Delta \rightarrow 0$ all amplitudes of the Fourier transformed idealized position eigenstate have the same magnitude, i.e., the spectrum is flat.

\subsection{Analysis}
\begin{figure}[!t]
\centering
\includegraphics[width=8.5cm]{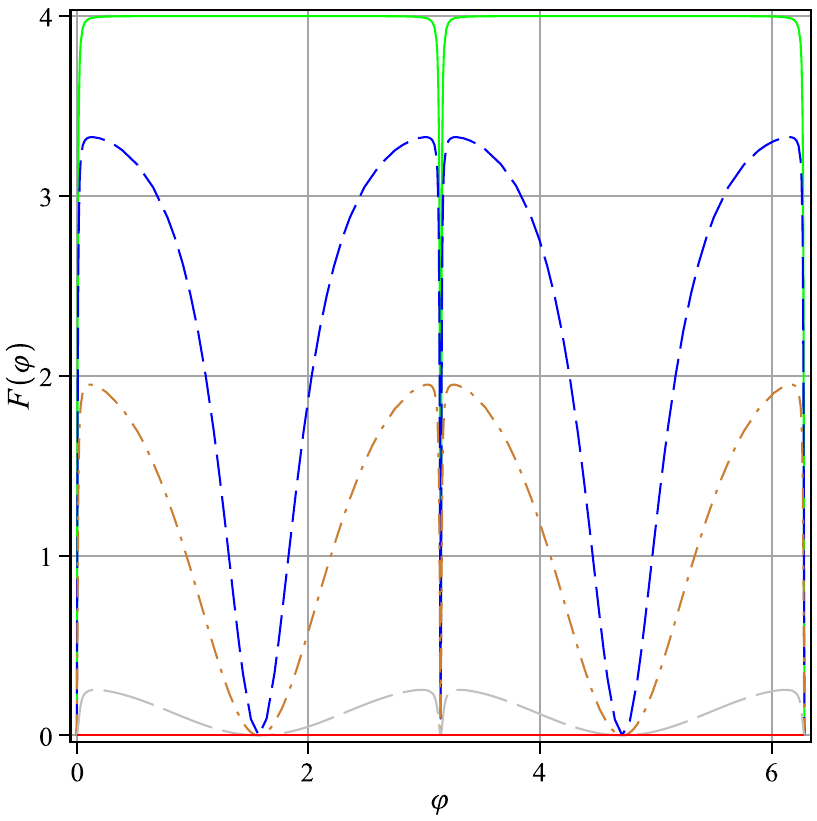}
\caption{General dependence of the Fisher information $F(\varphi)$ for five values of the parameter $r$: $r = 0$ corresponds to the uppermost solid line (green), $r = P/8$ corresponds to the dashed line (blue), $r = P/4$ corresponds to the dashed-dotted line (brown), $r = P/2$ corresponds to the long dashed line (gray), and $r = P$ corresponds to the lowermost solid line (red). \label{dep_r}}
\end{figure}

\noindent
Our procedure can be analyzed in two ways. As expected, from one perspective it behaves as a parameter estimation protocol. From the other, it  behaves as the Deutsch-Jozsa algorithm. First, we analyze the behavior of the parameter estimation part of the procedure. Based on the above measurement statistics, we calculate the Fisher information $F(\varphi)$. The minimal value of $F(\varphi) = 0$ occurs when function $f(x)$ is constant ($r = P$) with the corresponding measurement statistics
\begin{eqnarray}
p(x_{0}|\varphi) &=& \mbox{erf}^2(2 P \Delta), \nonumber \\
p(\bar{x}_{0}|\varphi) &=& 1 - \mbox{erf}^2(2 P \Delta). \nonumber
\end{eqnarray}
Conversely, the maximal value of the Fisher information
\begin{equation}
F(\varphi) = \frac{4 \, \mbox{erf}^2(2 P \Delta) (\cos(2 \varphi)-1)}{\mbox{erf}^2(2 P \Delta) (\cos(2 \varphi) + 1) - 2}
\end{equation}
occurs when function $f(x)$ is balanced ($r = 0$) with the corresponding measurement statistics
\begin{eqnarray}
p(x_{0}|\varphi) &=& \frac{1}{2} \, \mbox{erf}^2(2 P \Delta) \left[1 + \cos(2 \varphi)\right], \nonumber \\
p(\bar{x}_{0}|\varphi) &=& 1 - \frac{1}{2} \, \mbox{erf}^2(2 P \Delta) \left[1 + \cos(2 \varphi)\right]. \nonumber
\end{eqnarray}
Here, the optimal value of the Fisher information $F(\varphi) = 4$ is given for $\mbox{erf}^2(2 P \Delta) = 1 \Rightarrow P \geq 3/(2 \Delta)$ which, in general, implies $P \gtrsim 1/(2 \Delta)$ and is consistent with the approximation applied above. The general dependence of the Fisher information $F(\varphi)$ on parameter $r$ with $P = 3/(2 \Delta)$ and $\Delta = 1/\sqrt{2}$ (the variance of the coherent state) is shown in Fig.~\ref{dep_r}. The dips that are especially visible for the balanced function appear because the Fisher information $F(\varphi)$ retains some dependence on the parameter $\varphi$ since for $P = 3/(2 \Delta)$: $\mbox{erf}^2(2 P \Delta) \approx 1$.
Based on the general dependence of $F(\varphi)$ on $r$, we conclude that the maximal value of the Fisher information is indeed obtained for a balanced function.

To address the optimality of our parameter estimation protocol, we analyze the behavior of the generator of translations in the parameter $\varphi$: $\mathcal{H} \equiv f(\hat{x})$. The expectation value of the generator $\mathcal{H}$ in the state of the register CV preceding application of the black-box operator, i.e. $|\psi_{in}\rangle = F|r\rangle$ with $\Delta \ll T$, is given by
\begin{equation}
\langle \mathcal{H} \rangle = \langle f(\hat{x}) \rangle = \frac{1}{2} ( \mbox{erf}(2 P \Delta) - \mbox{erf}(2 r \Delta) ). \nonumber
\end{equation}
Since $f^2(x) = f(x)$ the variance of the generator $\mathcal{H}$ in $|\psi_{in}\rangle$ can be written as
\begin{eqnarray}
(\Delta \mathcal{H})^2 = (\Delta f(\hat{x}))^2 &=& \frac{1}{2} ( \mbox{erf}(2 P \Delta) - \mbox{erf}(2 r \Delta) ) \times \nonumber \\
&& \left[ 1 - \frac{1}{2} ( \mbox{erf}(2 P \Delta) - \mbox{erf}(2 r \Delta) ) \right]. \nonumber
\end{eqnarray}
The maximal expectation value of the generator $\mathcal{H}$ occurs for a balanced function ($r = 0$) with $P \geq 3/(2 \Delta)$ and is given by $\langle \mathcal{H} \rangle = 1/2$. On the other hand, the maximal variance of the generator $\mathcal{H}$ is $(\Delta \mathcal{H})^2 = 1/4$. Hence, the Fisher information is bounded by $F(\varphi) \leq 16(\Delta \mathcal{H})^2 = 4$. Therefore, we note that according to Eqs.~(\ref{bounds}) and (\ref{HL}) our procedure attains the scaling regime of the Heisenberg limit. However to establish its optimality we must calculate whether $\delta \varphi = 1/\sqrt{F(\varphi)}$. We use the standard expression for the mean squared error given by
\begin{equation}
\delta \varphi = \frac{\Delta X}{\left| d\langle X \rangle/d\varphi \right|}\, ,
\end{equation}
where $X$ is the measurement observable defined as $X = P_{x_{0}}$ [see Eq.~(\ref{POVM})]. Hence, for the final state $\ket{\psi_{\varphi}} = F^{-1} U_{f}(\varphi) F |r\rangle$ with $\varepsilon = \Delta$ we calculate $\langle X \rangle = \langle \psi_{\varphi}| P_{x_{0}} |\psi_{\varphi}\rangle = \frac{1}{2} \, \mbox{erf}^2(2 P \Delta) \left[1 + \cos(2 \varphi)\right]$.  Based on the property $P^{2}_{x_{0}} = P_{x_{0}}$ we find that $\langle X^{2} \rangle = \langle X \rangle$. For $P \geq 3/(2 \Delta)$ the mean squared error is $\delta \varphi = 1/2$. Hence, we conclude that for a balanced function our parameter estimation procedure over continuous variables attains the ultimate limit of the quantum Cram\'{e}r-Rao bound, and therefore is optimal. This result constitutes an analogy to the phase estimation with a qubit realized as a single photon placed in the arms of the Mach-Zender interferometer. Here, the balanced property of function $f(x)$ plays a role of two distinct paths in a balanced Mach-Zender interferometer.

Next, let us analyze the Deutsch-Jozsa side of the procedure. Under appropriate conditions the developed procedure can determine the character of function $f(x)$. If a value of the parameter $\varphi$ is fixed: $\varphi = \pi/2$ then the measurement statistics is given by
\begin{eqnarray}
p(x_{0}) &=& \mbox{erf}^{2}(2 r \Delta), \nonumber \\
p(\bar{x}_{0}) &=& 1 - \mbox{erf}^{2}(2 r \Delta), \nonumber
\end{eqnarray}
It is clear that for a constant and balanced function $f(x)$ the corresponding measurement statistics of the Deutsch-Jozsa algorithm are recovered. Indeed, when function $f(x)$ is constant ($r = P$) then
\begin{eqnarray}
p(x_{0}) &=& \mbox{erf}^{2}(2 P \Delta), \nonumber \\
p(\bar{x}_{0}) &=& 1 - \mbox{erf}^{2}(2 P \Delta), \nonumber
\end{eqnarray}
and when function $f(x)$ is balanced ($r = 0$) then $p(x_{0}) = 0$ and $p(\bar{x}_{0}) = 1$. The Deutsch-Jozsa algorithm over the semi-Gaussian states defined on a finite domain becomes a probabilistic procedure. This is consistent with the conclusions found in Ref.~\cite{sanders09}. However, when the size of the domain is sufficiently large with $P \geq 3/(2 \Delta)$ then a definite distinction between constant and balanced functions can be made. Nevertheless, even for large enough domains this implementation of the Deutsch-Jozsa protocol does not offer an unphysical, infinite speed-up over the classical procedures. We note that for ideal, nonnormalizable position eigenstates ($\Delta \rightarrow 0$), the constant function measurement statistics is retained for $P \rightarrow \infty$ rendering $P$ and $r$ unphysical, thus making a meaningful distinction between the balanced and constant functions impossible.

\begin{figure}[!h]
\centering
\includegraphics[width=8.5cm]{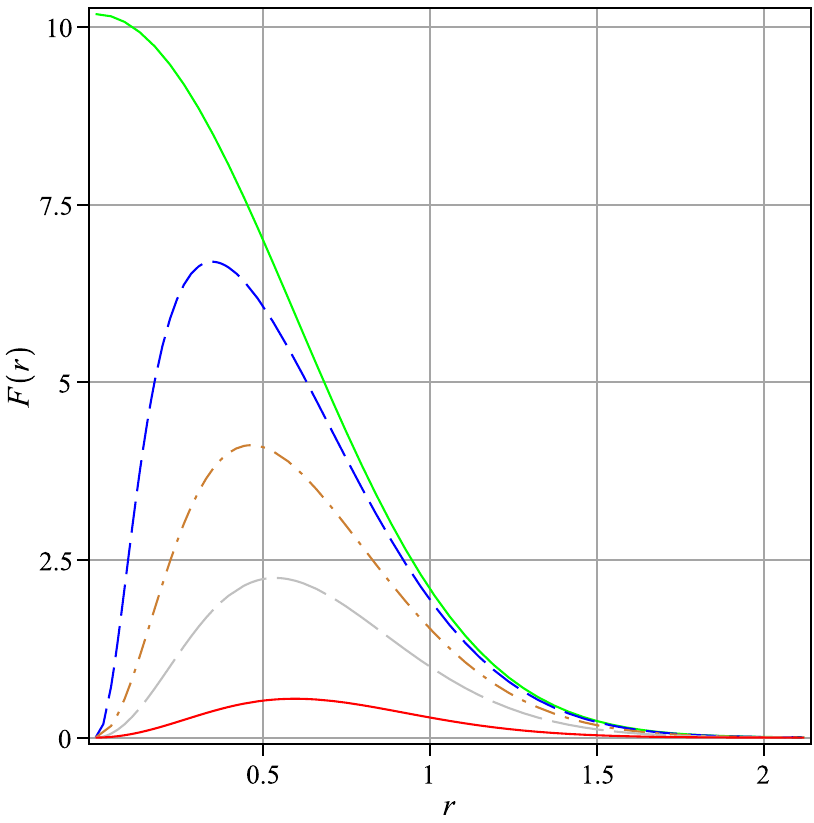}
\caption{General dependence of the Fisher information $F(r)$ for four values of the parameter $\varphi$: $\varphi = \pi/2$ corresponds to the uppermost solid line (green), $\varphi = 5 \pi/12$ corresponds to the dashed line (blue), $\varphi = \pi/3$ corresponds to the dashed-dotted line (brown), $\varphi = \pi/4$ corresponds to the long dashed line (gray), and $\varphi = \pi/8$ corresponds to the lowermost solid line (red). The optimal value of $r$ shifts from balanced to constant. \label{dep_phi}}
\end{figure}

We also calculated the Fisher information $F(r)$ and plotted it against $r \in (0, P)$ for five different values of the parameter $\varphi = \{\pi/2, 5 \pi/12, \pi/3, \pi/4, \pi/8\}$ with $P = 3/(2 \Delta)$ and $\Delta = 1/\sqrt{2}$ (see Fig.~\ref{dep_phi}).
The maximal value of the Fisher information $F(r)$ is obtained for $\varphi = \pi/2$ corresponding to a simplified Deutsch-Jozsa algorithm. We note that the optimality changes from balanced to more constant when $\varphi \neq \pi/2$. Any further analysis of this side of the procedure is problematic due to a lack of the generator of translations in $r$.

One possible application of the Deutsch-Jozsa part of our procedure is to test the quality of the implementation of function $f(x)$ employed in the parameter estimation protocol. Whenever the function is balanced or constant the quality of its implementation can be established by probing the parameter $r$. We also would like to stress that since we are employing a single continuous variable, no entanglement is present at the preparation stage nor is created during the computation. The quantum superposition principle itself is responsible for speed-up over any classical procedure. Even though, in principle, a single continuous variable is quite sufficient, a practical implementation of the Deutsch-Jozsa algorithm may require more continuous variables.

\section{Conclusions}\label{sec:conc}
\noindent
In conclusion, we developed a general procedure capable of performing two distinct tasks. For one mode of operation the protocol estimates a value of an unknown parameter with Heisenberg-limited precision. On the other hand, for a fixed value of the parameter in question the procedure addresses the Deutsch-Jozsa problem in a single run. Our procedure employs Fourier transforms and black-box unitary operator applied to a single continuous variable represented as the semi-Gaussian state defined on a finite domain. Consequently, for this setup, the parameter estimation side of the procedure is optimal and the Deutsch-Jozsa algorithm offers finite, i.e. physically feasible, speed-up over any classical procedure. Furthermore, no entanglement is present at any stage of the procedure. A similar conclusions concerning quantum metrology can be found in Refs.~\cite{tilma10,pinel10}. We emphasize a special role played by the balanced kind of functions $f(x)$. The procedure equipped with the black-box operator that introduces the parameter $\varphi$ via the balanced function attains the ultimate limit of the quantum Cram\'{e}r-Rao bound. This behavior can be linked to the phase estimation with a qubit realized as a single photon placed in the arms of the Mach-Zender interferometer.

\begin{acknowledgments}\noindent
This work was supported by the White Rose Foundation and the QIPIRC programme of the EPSRC.
\end{acknowledgments}

\end{document}